\documentclass[preprint,proceedings]{rmaa}

\usepackage{paralist}

\usepackage{psfrag,color}

\def\msun{$M_{\odot}$}
\def\mdot{$\dot M$}

\SetYear{2004}
\SetConfTitle{COMPACT BINARIES IN THE GALAXY AND BEYOND}

\title{SS433:the microquasar link with ULXs?} 

\author{
  P.A. Charles,\altaffilmark{1} 
  A.D. Barnes,\altaffilmark{1} 
  J. Casares,\altaffilmark{2} 
  J.S. Clark,\altaffilmark{3} 
  W.I. Clarkson,\altaffilmark{1}
  E.T Harlaftis,\altaffilmark{4}
  R.I. Hynes,\altaffilmark{5}
  T.R. Marsh,\altaffilmark{1}
and  D. Steeghs,\altaffilmark{6}}

\altaffiltext{1}{University of Southampton, Southampton, UK}
\altaffiltext{2}{Instituto de Astrof\'\i{}sica de Canarias, La Laguna, Spain.}
\altaffiltext{3}{University College London, UK}
\altaffiltext{4}{National Observatory, Athens, Greece}
\altaffiltext{5}{University of Texas at Austin, USA}
\altaffiltext{6}{Harvard-Smithsonian Center for Astrophysics, Cambridge, USA.}

\shortauthor{P.A. Charles et al.}
\shorttitle{SS433:the microquasar link with ULXs?}

\fulladdresses{
\item A. Barnes, P.A. Charles: School of Physics \& Astronomy,
University of Southampton, Southampton, SO17 1BJ, UK
  (\email{adb@astro.soton.ac.uk; pac@astro.soton.ac.uk}).
\item J. Casares: Instituto de Astrof\'\i{}sica de Canarias, 38200 La Laguna, 
Tenerife, Spain 
  (\email{jcv@ll.iac.es}).
\item J.S. Clark: Dept of Physics \& Astronomy, University College
London, Gower Street, London, WC1, UK 
  (\email{jsc@}).
\item W.I. Clarkson: Dept of Physics \& Astronomy, Open University, Milton Keynes, MK7 6AA, UK
  (\email{W.I.Clarkson@open.ac.uk}).
\item E.T. Harlaftis:Institute of Space Applications and Remote Sensing,
National Observatory of Athens, P.O. Box 20048,
Thessio, Athens - 118 10, Greece
  (\email{ehh@space.noa.gr}).
\item R.I. Hynes: University of Texas at Austin, Astronomy Department, 
1 University Station C1400, Austin, Texas 78712, USA
  (\email{rih@obelix.as.utexas.edu}). 
\item T.R. Marsh: Dept of Physics, University of Warwick, Coventry CV4 7AL, UK
  (\email{t.marsh@warwick.ac.uk}).
\item D. Steeghs: Harvard-Smithsonian Center for Astrophysics, 60 Garden 
Street, MS-67, Cambridge, MA 02138, USA 
  (\email{dsteeghs@cfa.harvard.edu}).

}

\listofauthors{P.A. Charles, A.D. Barnes, J. Casares, J.S. Clark,
W.I. Clarkson, E.T. Harlaftis, R.I. Hynes, T.R. Marsh, \& D. Steeghs}

\indexauthor{Charles, P.A.}
\indexauthor{Barnes, A.D.}
\indexauthor{Casares, J.}
\indexauthor{Clark, J.S.}
\indexauthor{Clarkson, W.I.}
\indexauthor{Harlaftis, E.T.}
\indexauthor{Hynes, R.I.}
\indexauthor{Marsh, T.R.}
\indexauthor{Steeghs, D.}

\abstract{SS433 is the prototype {\it microquasar} in the Galaxy and
may even be analogous to the ULX sources if the jets' kinetic energy
is taken into account.  However, in spite of 20 years of study, our
constraints on the nature of the binary system are extremely limited
as a result of the difficulty of locating spectral features that can
reveal the nature and motion of the mass donor. Newly acquired, high
resolution blue spectra taken when the (precessing) disc is edge-on
suggest that the binary is close to a common-envelope phase, and hence
providing kinematic constraints is extremely difficult.  Nevertheless,
we do find evidence for a massive donor, as expected for the inferred
very high mass transfer rate, and we compare SS433's properties with
those of Cyg X-3.}

\addkeyword{accretion}
\addkeyword{accretion discs}
\addkeyword{X-rays: binaries}
\addkeyword{X-ray: stars}

\begin{document}
\maketitle

\section{Introduction}
\label{sec:intro}
SS433 holds a special place in late 20th century astronomy as the
first relativistic jet source discovered in the Galaxy (see Margon
1984), and hence is the prototype {\it microquasar}.  It is remarkable
as the only {\it continuously} emitting microquasar, with its key
feature being the 162.5d precession period of the jets (and associated
accretion disc) that is revealed by the ``moving lines'' which have
now been observed for more than 2 decades.  The radial velocity curves
of the moving lines are very well described by the {\it Kinematic
Model} (Margon 1984), thereby revealing key parameters of the jets
($v$=0.26c; $i$=79$^\circ$) and, combined with radio observations of
the associated SNR W50, a distance of 4.8kpc.  It is observed as a
weak ($\sim$10$^{36}$erg~s$^{-1}$) X-ray source (X1909+048), which
combined with the high orbital inclination (it is eclipsing) suggests
that it may be an Accretion Disc Corona (ADC) type of X-ray source.
However, it is well established that the kinetic energy in the jets is
at least 1,000 times greater than the observed luminosity (Safi-Harb
2003), although the origin of the jet energy is still a matter of
debate, and this would place SS433 in the recently defined class of
ULX sources (e.g. King 2003). If the intrinsic luminosity were indeed
$\geq$10$^{39}$erg~s$^{-1}$, then the implied mass-transfer rate would
be extremely high (\mdot$\sim$10$^{-4}$\msun$y^{-1}$, see also Fuchs et al
2003).

This has led to King et al (2000) investigating the evolutionary
status of SS433 and they show that, given a massive, radiative
companion, it is possible for SS433 to avoid the common-envelope
phase, by expelling the majority of the transferred material (part of
which constitutes the jets).  They conclude that the observed X-rays
are then likely to almost entirely arise from the jets themselves.
Indeed, the X-ray output of SS433 appears to be dominated by the
precessing jets, with moving X-ray emission lines and even spatially
resolved X-ray structure along the jet axis (Marshall et al 2002;
Namiki et al 2003). Recently, however, the RXTE All-Sky Monitor has
revealed a long-term modulation in the precessional X-ray lightcurve
(Gies et al 2001). The relativistic doppler boosting from jet
precession alone is insufficient to produce this modulation, which
requires an additional, presumed geometrical component (Clarkson et al
2004).  For the established high inclination this can most naturally
be provided by structure in the edge-on view of the accretion disk,
which may be tilted or warped to produce the apparent asymmetry in the
precessional lightcurve.

\begin{figure}[!t]\centering
  \vspace{0pt}
  \includegraphics[angle=0,width=0.94\columnwidth]{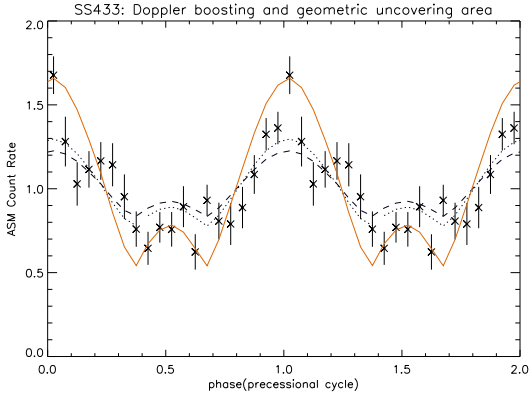}
  \caption{7-year X-ray lightcurve of SS 433
obtained with the RXTE/ASM. The dotted and dashed curves represent
modulation expected from doppler boosting in the relativistic jets (for
the cases of bullets and a continuous stream respectively). An additional
geometric modulation is clearly required, modelled here as the modulation
of a power-law jet by a precessing accretion disk (dot-dashed line). From
Clarkson et al (2004).}
  \label{fig:asm}
\end{figure}

\section{Dynamical constraints}
\label{sec:dynamical}

While the binary period (of 13d) is also well-known, the
nature of the mass donor (a $\sim$B star; Crampton \& Hutchings 1981)
is poorly constrained because of the high reddening in combination
with strong disc/jet emission features that obliterate the
characteristic (and weak) absorption lines of early-type stars.
Consequently, only emission line radial velocity curves have been
derived with any confidence (Fabrika \& Bychkova 1990), which means
that dynamical constraints on the system masses are limited.  Indeed,
with no evidence for X-ray pulsations or bursts, there is no obvious
indication as to the nature of the compact object.  And the
interpretation of the radial velocity curves requires assumptions
about the mass donor, leading to estimates
that range from 0.8M$_\odot$ (d'Odorico et al 1991) to 62M$_\odot$
(Antokhina \& Cherepashchuk 1985)!

Nevertheless Fabrika \& Bychkova used $\sim$4 yrs of 6m data taken
around precession phase $\Psi\sim$0 (which minimises the scatter in their radial
velocity curve), corresponding to the most open viewing angle of the disc
(and hence best view of the hotter inner disc regions).  For an assumed $e$=0
they obtain $f(M_2)\simeq$8\msun, which gives $M_2\sim$10\msun~for a NS,
or 15\msun~for a 6\msun~BH (both are capable of explaining the
extremely high \mdot).

Consequently it has been impossible to add SS433 to the growing list
of compact object mass determinations (Charles \& Coe 2003) which is
so important for a system whose properties link it to both the
microquasars and ULX systems.

\section{Detection of mass donor in SS433?}
\label{sec:donor}

\begin{figure}[!t]
  \includegraphics[width=\columnwidth, height=6.8cm]{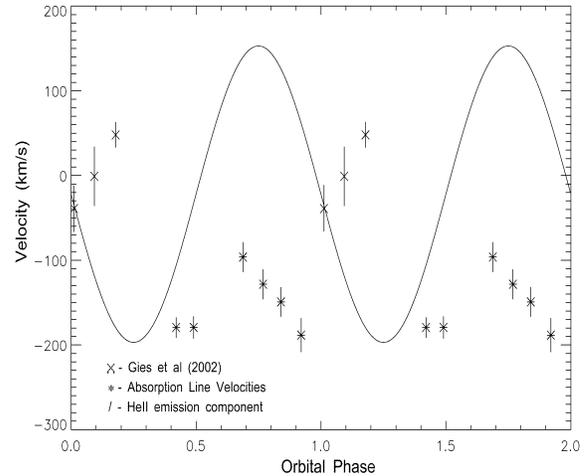}
  \caption{Absorption line radial velocity measurements of SS433
  obtained by us (*) and Gies et al 2002 (x), together with the
  Fabrika \& Bychkova 1990 fit to their HeII emission velocities which were
  presumed to represent the disc/compact object.}
  \label{fig:RV}
\end{figure}

Recognising the need to investigate the nature of the mass donor if
progress was to be made in understanding SS433, it became clear to
several groups that high S/N, high resolution blue spectra were crucial
in order to search for weaker spectral features in absorption (away from the strong
disc emission lines).  Assuming that the best chance to search for the
donor signature would come when the disc had precessed (with the jets)
to its maximum opening angle, Gies et al (2002) obtained spectra that
suggested weak absorption features similar to those seen in late A
supergiants.  These did yield a few radial velocity measurements, but could
only be used to derive mass constraints in combination with the HeII
emission radial velocity curve (see figure~\ref{fig:RV}), from which
they obtained $q=M_X/M_2=0.6\pm0.1$, and $M_X=11\pm5$M$_\odot$,
$M_2=19\pm7$M$_\odot$.  Such an extreme mass ratio would be expected
in order to explain the required very high (thermal) mass transfer
rate (King et al 2000). However, their extremely limited orbital phase coverage
clearly indicated the need for further observations to confirm and
extend these results.

\begin{figure}[!t]\centering
  \vspace{0pt}
  \includegraphics[angle=-90,width=0.94\columnwidth]{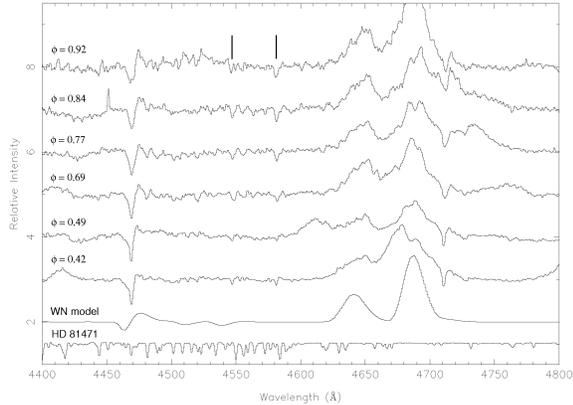}
  \caption{Blue spectra of SS433 obtained near precession phase
  $\sim$0.5, when the disc is closest to edge on.  The sharp
  absorption features between $\lambda\lambda$4500-4600 are
  characteristic of late B--A spectral types and we associate these
  with the mass donor.  The bottom two spectra show (upper) the
  similarity of the gross SS433 emission features to that of a high
  mass-loss WN star and (lower) an A7I template star HD 81471. }
  \label{fig:spectra}
\end{figure}

In fact, we had been acquiring such blue spectra of SS433 for several
years (from AAT, WHT, INT, and Calar Alto), in an attempt to obtain
good sampling in both orbital and precessional phases.  Our initial
aim was to search for any irradiation signature of the donor via the
Bowen fluorescence mechanism, as has been so successful in Sco X-1 and
other luminous X-ray binaries (see Casares et al, these proceedings).
Unfortunately, no such component was visible, only broad Bowen
emission, likely from an extended region such as the disc.  We had
also decided to take a different route to Gies et al when seeking
evidence for the donor itself, and therefore examined our spectra
obtained around precessional phases $\sim$0.3--0.7, when the jets are
close to perpendicular to our line-of-sight, and correspondingly the
accretion disc is seen close to edge-on.  Since the strong optical
variability as a function of precessional phase is already
well-established (and indicates a disc contribution at least equal to
that of the mass donor), we felt that it was important to minimise the
disc component by observing when the disc was edge-on.  Our best
spectra obtained at these phases are shown in
figure~\ref{fig:spectra}.  Of particular interest are the weak, but
narrow, absorption features in the $\lambda\lambda$4500-4600 region.
These are mostly FeII transitions and are typically seen in early-type
(late B - A) spectra.  This includes the spectral type suggested by
Gies et al, but we could not detect the transitions they reported.
However, we have now clearly detected such features in SS433, and
these are marked in figure~\ref{fig:spectra}, and are strong in the A7
template (HD81471) plotted at the bottom.

What was more remarkable was the radial velocity curve of these
absorption features (see figure~\ref{fig:RV}).  While the velocity
around orbital phase 0.75 is close to that expected from Gies et al's
values for their mass ratio, the subsequent velocities (approaching
phase 1.0) move in the {\it opposite} direction!  This only makes
sense if the secondary is embedded in a dense outflowing disc wind,
although we might expect these lines to then show P Cyg behaviour
under such circumstances, and the velocities around phase 0.4--0.5 are
even more anomalous.  Nevertheless, with this orbital phase behaviour,
it becomes extremely difficult to justify a keplerian interpretation
of the observed velocities, and hence any resulting masses would be
subject to major systematic effects.  And support for the concept of a
major disc plane (equatorial) mass outflow has already been obtained
via the radio detection of extended emission in a direction that is
{\it perpendicular} to the main jet outflow (Blundell et al 2001).
Furthermore, the overall SS433 spectrum is remarkably similar to
models of WN spectra (e.g. 2nd from bottom plot of
figure~\ref{fig:spectra} from Clark \& Porter 2004 which is a model
for the WN9 star WR105 with parameters of $T$ = 32,500K, \mdot =
1.2$\times$10$^{-5}$\msun$y^{-1}$ and $v_\infty$ = 700~km~s$^{-1}$),
and represents a dense outflow in which an ionising source is
embedded.  We are not claiming that SS433 is a W-R star, but simply
point out that it makes little difference whether a hot star or
powerful X-ray source is the source of that ionisation, both of which
are of course present in Cyg X-3 (Fuchs et al 2003), which may even be
the end result of the evolution of systems like SS433.


\begin{thebibliography}

\bibitem{} Blundell, K.M. et al. 2001, ApJ, 562, L79
\bibitem{} Casares, J. et al. 2003, ApJ, 590, 1041
\bibitem{} Charles, P. \& Coe, M.J. 2003, astro-ph/0308020
\bibitem{} Clark, J.S. \& Porter, J. 2004, A\&A submitted
\bibitem{} Clarkson, W.I. et al. 2004 (in preparation)
\bibitem{} Crampton, D. \& Hutchings, J.B. 1981, ApJ, 251, 604
\bibitem{} Fabrika, S.N. \& Bychkova, L.V. 1990, A\&A, 240, L5
\bibitem{} Fuchs, Y. et al. 2003, astro-ph/0208432
\bibitem{} Gies, D.R. et al. 2002, ApJ, 578, L67
\bibitem{} Hynes, R.I. et al. 2003, ApJ, 583, L95
\bibitem{} King, A.R. 2003, astro-ph/0301118
\bibitem{} Marshall, H.L. et al. 2002, ApJ, 564, 941 
\bibitem{} Namiki, M. et al. 2003, PASJ, 55, 281
\bibitem{} Safi-Harb, S. 2003, in {\it New Views on Microquasars}, p243
\bibitem{} Steeghs, D. \& Casares, J. 2002, ApJ, 568, 273
\bibitem{} Wade, R.A.. \& Horne, K. 1988, ApJ, 324, 411

\end{thebibliography}
\end{document}